\documentclass[preprint,superscriptaddress,prd,showpacs,aps]{revtex4-1}
\usepackage{txfonts}
\usepackage{wasysym}
\usepackage{graphicx,epsfig,amssymb}
\usepackage{multirow}
\usepackage{verbatim}
\usepackage{color} 

\newcommand{\newc}{\newcommand}	
\newc{\gsim}{\lower.7ex\hbox{$\;\stackrel{\textstyle>}{\sim}\;$}}
\newc{\lsim}{\lower.7ex\hbox{$\;\stackrel{\textstyle<}{\sim}\;$}}

\def\beq{\begin{equation}}
\def\eeq{\end{equation}}
\def\beqn{\begin{eqnarray}}
\def\eeqn{\end{eqnarray}}

\def\tth{t\bar t H}

\begin{document}
\title{Associated Production of Higgs Boson and $t\bar t$ at LHC}

\author{Hong-Lei Li}\email{sps$_$lihl@ujn.edu.cn}
\affiliation{School of Physics and Technology, University of Jinan, Jinan Shandong 250022,  China}

\author{Peng-Cheng Lu}\email{l$_$pc221@163.com}
\affiliation{School of Physics, Shandong University, Jinan, Shandong 250100,  China}
\affiliation{School of Physics and Technology, University of Jinan, Jinan Shandong 250022,  China}

\author{Zong-Guo Si}\email{zgsi@sdu.edu.cn}
\affiliation{School of Physics, Shandong University, Jinan, Shandong 250100,  China}
\affiliation{CCEPP,IHEP,Beijing 100049, China}

\author{Ying Wang}\email{wang$_$y@mail.sdu.edu.cn}
\affiliation{School of Physics, Shandong University, Jinan, Shandong 250100,  China}

\begin{abstract}
One of the future goals of the LHC is to precisely measure the properties of Higgs boson. 
The associated production of Higgs boson and the top quark pair is a promising process to 
investigate the related Yukawa interaction and the properties of Higgs. Compared with the pure 
scalar sector in the Standard Model, the Higgs sector contains both scalar and pseudoscalar 
in many new physics models, which makes the $\tth$ interaction more complex and provides 
a variety of phenomena. To investigate the $\tth$ interaction and the properties of Higgs, 
we study the top quark spin correlation observables at the LHC.
\end{abstract}
\maketitle	
%%%%%%%%%%%%%%%%%%%%%%%%%%%%%%%%%%%%%%%%%%%%%%%%%%%%%%%%%%%%%%%%%%%%%%%%%
\section{Introduction}
%%%%%%%%%%%%%%%%%%%%%%%%%%%%%%%%%%%%%%%%%%%%%%%%%%%%%%%%%%%%%%%%%%%%%%%%%
The discovery of Higgs boson, confirmed by the ATLAS and CMS experiments at the 
LHC~\cite{ATLAS-Higgs,CMS-Higgs}, provides the forceful evidence for the 
Brout-Englert-Higgs mechanism in the 
Standard Model (SM). The impressive accurate experimental results not only 
support the success of the SM but also push the theoretical predictions forward 
to a higher accuracy. Except for the mass and spin, other properties of Higgs 
boson should be clear to achieve the completeness of the SM. It is well known 
that the masses of fermion are extracted from the related Yukawa interactions, which 
offer the opportunity to study the interaction of Higgs boson and fermions. 
Due to the large quark mass, the Yukawa coupling of $\tth$ is order of one. 
Therefore a large number of phenomena can be studied in the Higgs boson production 
associated with the top quark pair. 

The precise measurements on the Higgs sector are indispensable for the understanding 
of the origin of electroweak symmetry breaking. Latest results on the Higgs mass as 
well as the spin and parity~\cite{ATLAS_mass2,CMS_mass_coup,ATLAS_spin,CMS_spin} are 
reported by the ATLAS and CMS collaborations. At the same time the couplings of Higgs
 boson are consistent with the predictions in the SM~\cite{ATLAS_coupling,CMS_mass_coup}. 
However, it is possible for the observed Higgs boson to be one scalar sector in the other 
physics models, such as the Two Higgs Doublet Models~\cite{TDLEE,SW,Liu:1987ng}, the 
Minimal Supersymmetric Models~\cite{Wess:1973kz,Haber:1984rc,Martin:1997ns} and the 
Left-right symmetric Models~\cite{Pati:1974yy,Mohapatra:1974hk,Mohapatra:1974gc,
Senjanovic:1975rk,Mohapatra:1977mj}, etc. Top quark, as the known heaviest quark, 
is expected to decay before hadronization for its short
 life time. Hence, its spin property can be transferred to its decay products. The spin effects 
of $t\bar t$ production have been studied at the hadron colliders 
~\cite{Bernreuther:2004wz,Bernreuther:2010ny,Bernreuther:2012sx,AguilarSaavedra:2012rx,
Bernreuther:2004jv,Bernreuther:2000yn,Bernreuther:2002rc,Brandenburg:2002xr,Bernreuther:2001rq,
Bernreuther:2001bx}. It is found that the
top quark spin effects are sensitive to the new interactions~\cite{Gopalakrishna:2010xm,Gong:2012ru,
Bernreuther:2013aga}.
Investigating the $\tth$ production is helpful
to discriminate the Higgs boson among various models. The leading-order and the next-leading-order 
$\tth$ cross section predictions have been accomplished in the literatures~\cite{Kunszt:1984ri,Marciano:1991qq,Dawson:2002tg,Beenakker:2001rj,Beenakker:2002nc,Dawson:1997im,
Dittmaier:2011ti}. Both the ATLAS and CMS experiments have performed the measurement on the 
$\tth$ production with $H\to b\bar b$ at $\sqrt{S}= 8$ TeV~\cite{ATLAS_tth,CMS_tth}. Some 
phenomena on $\tth$ interaction have been studied in ~\cite{Curtin:2013zua,CMS:2013fda,
Nishiwaki:2013cma,Maltoni:2002jr,Biswas:2014hwa,He:2014xla}.
The reconstruction of $\tth$ signal 
and the corresponding backgrounds analysis have been studied in details~\cite{Buckley:2013auc,Santos:2015dja}.
The CP-properties of $\tth$ interaction play an important roles in the understanding of Yukawa interaction, which 
can be probed through the Higgs production in association with the top quark pair at the LHC~\cite{Frederix:2011zi,Ellis:2013yxa,Demartin:2014fia,Boudjema:2015nda}. 
 Besides the topics discussed in the previous works, in this paper we concentrate on the 
spin observables with different Higgs mass for scalar. Then we discuss  
systematically the spin observables for the scalar, pseudoscalar and mixing 
Higgs in $\tth$ production at the LHC. Additionally, we also simply 
investigate the corresponding background processes for our signal process 
$p p \to t \bar t H \to b \bar b l^+ l^- \nu_l \bar {\nu_l} + b \bar b$. These results both at LHC 13 TeV and 33TeV  
can help to study the Higgs properties.

This paper is organized as follows. The $\tth$ interactions in the SM and the Two Higgs 
Doublet Models are reviewed in Section II, together  with the brief introduction of top quark spin 
correlations. After that the $\tth$ production at the LHC and the corresponding observables are analyzed
in Section III. Finally,  a short summary is given.

%%%%%%%%%%%%%%%%%%%%%%%%%%%%%%%%%%%%%%%%%%%%%%%%%%%%%%%%%%%%%%%%%%%%%%%%%
\section{The $\tth$ interaction and the top quark spin effects}
%%%%%%%%%%%%%%%%%%%%%%%%%%%%%%%%%%%%%%%%%%%%%%%%%%%%%%%%%%%%%%%%%%%%%%%%%
In the SM, the scalar sector includes one SU(2) doublet. After spontaneous symmetry breaking, the 
Higgs boson couples to the top quark in the formula of 
\begin{equation}
{\cal L}_{t\bar t H}(SM)=-\frac{m_t}{v}t\bar  t H,
\end{equation}
where $m_t$ is the top quark mass and $v$ is the vacuum~expectation value of Higgs. 
Naturally, this interaction is CP-even under CP transformation. 

Compared with the single SU(2) doublet in the SM, it includes an additional SU(2) doublet in
 the two-Higgs-Doublet Model for the scalar sector. It means there are two vacuum expectation 
values, $v_1$ and $v_2$. Therefore, one generalized representation for the scalar doublet is 
\begin{equation}
\Phi_\alpha=\left(\begin{array}{c}
\phi^{+}_{\alpha} \\ (v_\alpha+\rho_\alpha+i\eta_\alpha)/\sqrt{2} \end{array}
\right),~~~~~~~~~(\alpha=1,2),
\end{equation}
with $v_1=v\cos\beta$ and $v_2=v\sin\beta$ and $\phi^{+}_\alpha$, $\rho_\alpha$ and $\eta_\alpha$ 
are the scalar fields. After the spontaneous symmetry breaking, 
there remain five physical Higgs particles: two 
CP-even $H_1$ and $H_2$ bosons, one CP-odd $A$ boson, and two charged $H^{\pm}$ bosons. Naturally,
 the light neutral Higgs $H_1$ could be regard as the SM Higgs boson. Following 
that the Yukawa coupling for the neutral Higgs bosons can be obtained from the Lagrangian
\begin{equation}
{\cal L}_{f\bar f H}=-\sum_{f=u,d,l}\frac{m_f}{v}(\epsilon^f_{H_1}f \bar f H_1 +
\epsilon^f_{H_2}f \bar f H_2-i\epsilon^f_{A}f \gamma_5\bar f A),
\label{eq:lag2hdm}
\end{equation}
where it brings three types of Yukawa interaction for the SM-like Higgs, the heavy Higgs and 
the pseudoscalar Higgs. Hence, the $\tth$ interaction gets more complicated and leads to different
properties for the production. 

For the top quark lifetime is extremely short, it decays without hadronization. Thus the decay 
production  becomes important to analyze the top quark spin information. Defining the angle of decay 
particle's ($f$'s) moving direction with the top quark spin polarization direction as $\theta$, one can obtain the 
distribution of the decay production in the formula of
\begin{equation}
\frac{d\sigma}{\sigma d \cos \theta}=\frac{1}{2}\left( 1+p c_f \cos \theta \right),
\end{equation} 
where $p$ is the polarization degree, $c_f$ is the spin-analyzer power of $f$, and $f$ can be $l^+$, 
$\nu$, $W^+$, $q$, $\bar q$.  In the SM, the tree level result shows that $c_{l^+}=c_{\bar d}=1$,
$c_{\nu}=c_u=-0.3$, and $c_b=c_{W^+}=-0.39$. Thus the charged lepton and the down type light quark are the best spin
 analyzers.  
 
In the top quark pair production, the spin correlation is related to the final charged lepton. 
According to the spin correlation,
the spin asymmetry of $t\bar  t$ manifests in the decay particle double distribution
\begin{eqnarray}
    {1\over \sigma}{d\sigma\over d\cos\theta_1 d\cos\theta_2}=  {1\over 4} (1 + {A}_1 \cos\theta_1  
+ {A}_2 \cos\theta_2 - {A}_3 \cos\theta_1 \cos\theta_2)\,\, ,
 \label{eq:ddist}
\end{eqnarray}
where $\sigma$ denotes the cross section of the respective reaction.  Here  $\theta_1$ ($\theta_2$) is
 the angle between the direction of flight of the lepton $\ell^+$ or jet $j_1$ ($\ell\,'^-$ or $j_2$) 
from $t$ $({\bar t})$ decay in the $t$ ($\bar{t}$) rest frame  and various polarization directions. 
The coefficient $A_1$($A_2$) reflects the single spin effects in $t\bar t$ production and $A_3$ is 
a measure of $t \bar t$ spin correlations. The polarization and the spin correlation provides
 important information on the dynamics of the top quark. 
The similar distributions of $\tth$ production at the hadron colliders can be obtained, which support to
investigate the spin effects in $\tth$ production at the LHC.
%%%%%%%%%%%%%%%%%%%%%%%%%%%%%%%%%%%%%%%%%%%%%%%%%%%%%%%%%%%%%%%%%%%%%%%%%
\section{The Phenomenology of $\tth$ Production at the LHC}
\begin{figure}
\centering 
\includegraphics[width=0.30\textwidth]{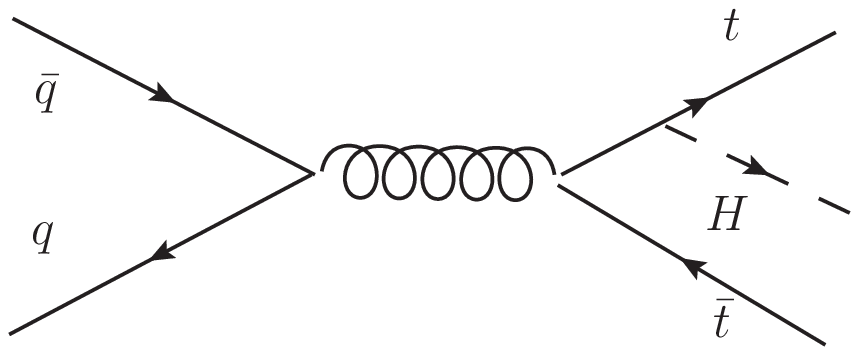}
\includegraphics[width=0.30\textwidth]{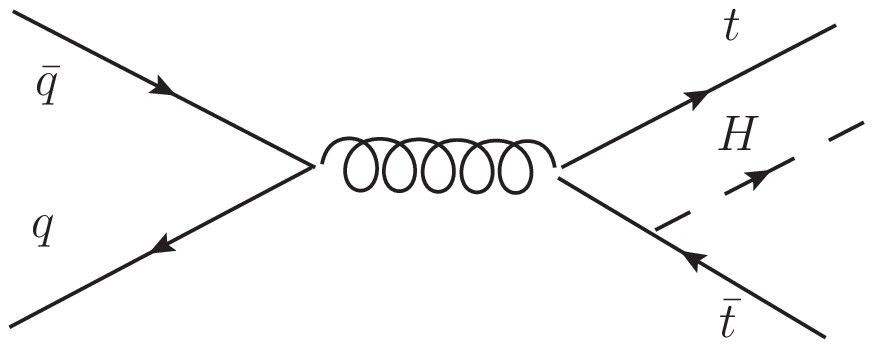} \\ \vspace{3 mm}
\includegraphics[width=0.30\textwidth]{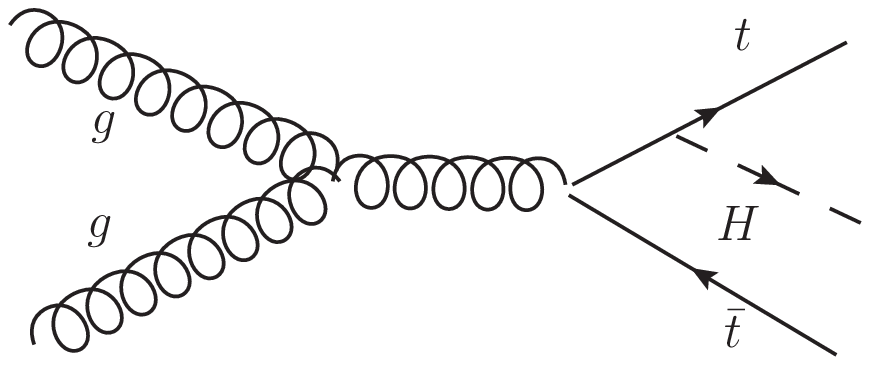}
\includegraphics[width=0.30\textwidth]{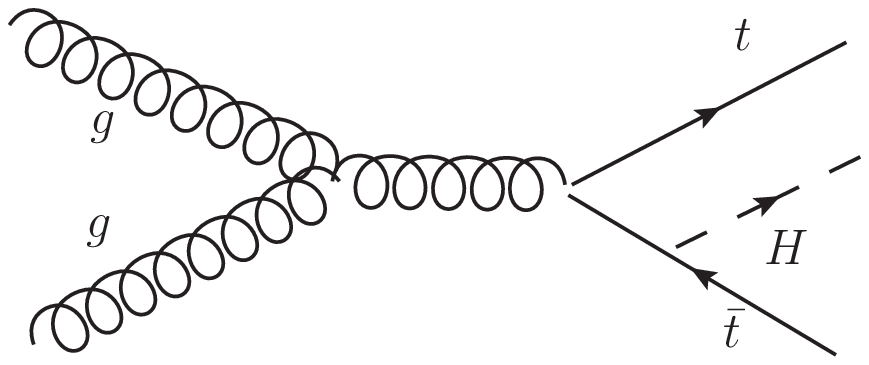} \\ \vspace{5 mm}
\includegraphics[width=0.30\textwidth]{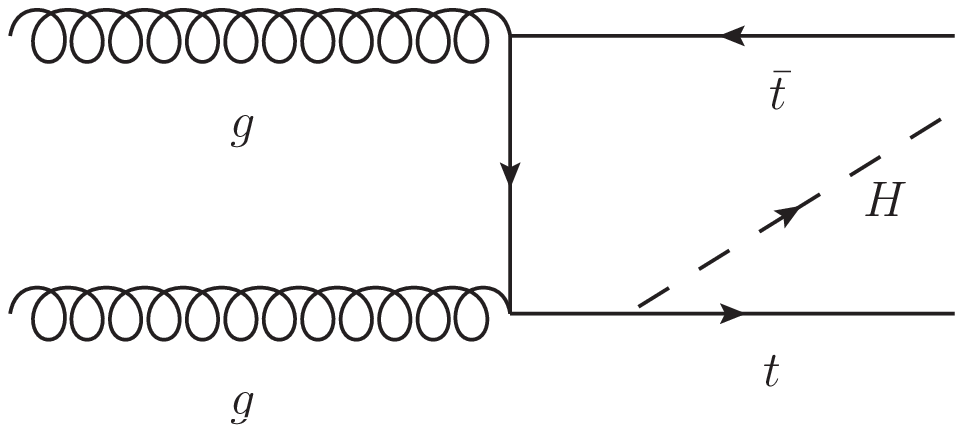}
\includegraphics[width=0.30\textwidth]{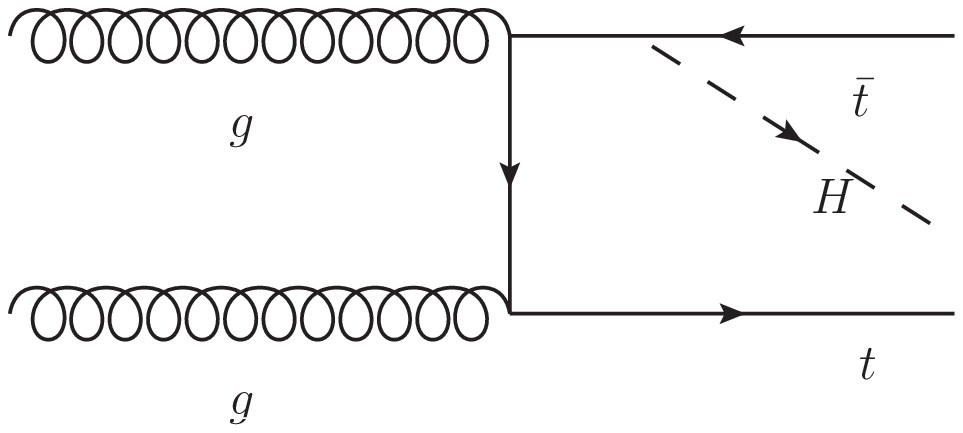}
\includegraphics[width=0.30\textwidth]{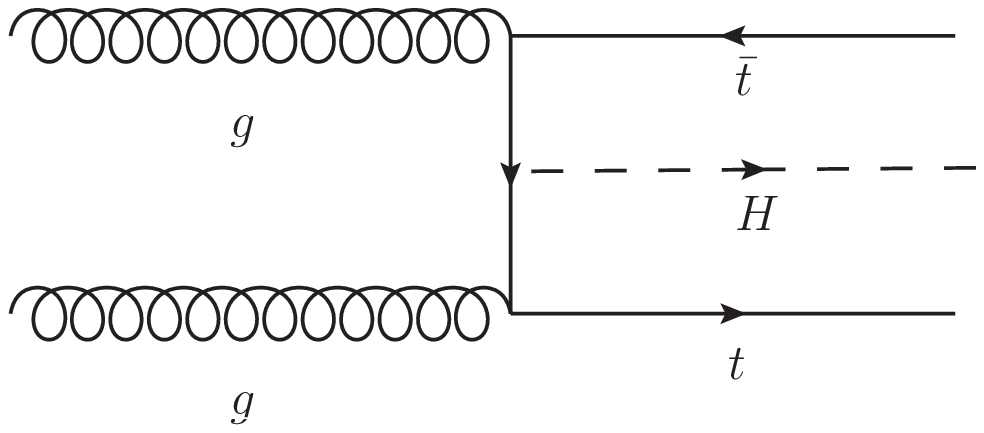}
\caption{The Feynman diagrams for $\tth$ production at the leading order QCD.}
\label{fig:feydiag}
\end{figure}

%%%%%%%%%%%%%%%%%%%%%%%%%%%%%%%%%%%%%%%%%%%%%%%%%%%%%%%%%%%%%%%%%%%%%%%%%
At the proton-proton colliders the $\tth$ is produced via the quark annihilation and the gluon fusion processes, which 
are displayed in Fig.~\ref{fig:feydiag} at the leading order QCD. Due to the charged lepton is a good 
trigger for the detector at the proton-proton collider (e.g. LHC), we investigate the $\tth$ production process with 
both top quarks leptonic decay,
\begin{equation}
pp\to t\bar{t}H \to b l^+ \nu \bar{b} l^- \nu H, ~~~~~~(l=e,\mu).
\label{pp2tth}
\end{equation}

One can notice that the $\tth$ interaction is different for scalar and pseudoscalar Higgs from equation (\ref{eq:lag2hdm}). In the follows, 
we respectively study the $t \bar t$ production in association with light Higgs (SM-Higgs), Heavy Higgs and 
scalar-pseudoscalar mixing Higgs. The top quark mass is set at 173.2 GeV and the SM-Higgs mass is at 125 GeV 
for the numerical results. 

%%%%%%%%%%%%%%%%%%%%%%%%%%%%%%%%%%%%%%%%%%%%%%%%%%%%%%%%%%%%%%%%%%%%%%%%%
\subsection{The SM-Higgs production associated with $t \bar t$}
\begin{figure}
\centering
\includegraphics[width=0.55\textwidth]{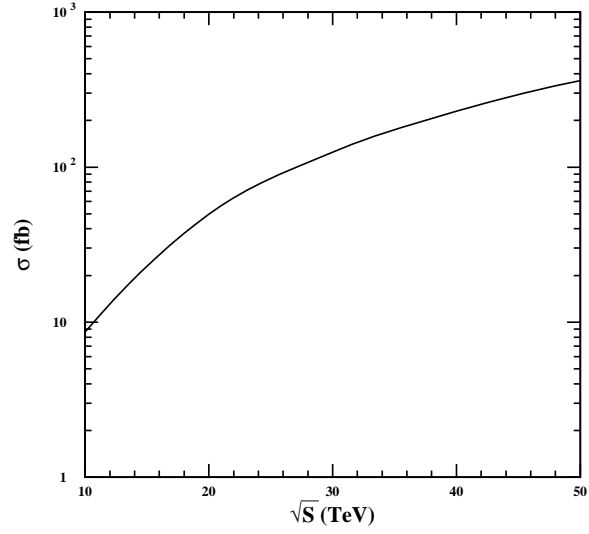}
\caption{The cross section for process (\ref{pp2tth}) with respect to the center-of-mass energy for SM-Higgs.}
\label{fig:cs-sm}
\end{figure}

%%%%%%%%%%%%%%%%%%%%%%%%%%%%%%%%%%%%%%%%%%%%%%%%%%%%%%%%%%%%%%%%%%%%%%%%% 
The cross section corresponding to process (\ref{pp2tth}) with different center-of-mass energy at 
the proton-proton collider
is plotted in Fig.~\ref{fig:cs-sm}. The event number for $\tth$ production with leptonic top quark 
decay is about 5000 (17000) at  LHC 13 TeV with the integrated luminosity of $300$($1000$) $\rm{fb}^{-1}$,
which can be up to 20000 (150000) at LHC 14 (33) TeV with a integrated luminosity of $1000$ $\rm{fb}^{-1}$.
It refers that the cross section can be tuned by a K-factor of 1.2-1.5 
from the high order calculations~\cite{Dawson:2002tg,Beenakker:2001rj,Beenakker:2002nc,
Dawson:1997im,Dittmaier:2011ti}.

According to equation (\ref{eq:ddist}), the coupling of $\tth$ is related to the distribution of the final 
state leptons decayed from the (anti-)top quark. Based on the  $t \bar t$ spin correlations, the 
observables related to the $\tth$ interaction can be defined as
\begin{eqnarray*}
O_1&=&\hat{q}_+\cdot \hat{q}_-, \\
O_2&=&(\hat{q}_+\cdot \hat{k}_t) (\hat{q}_-\cdot \hat{k}_{\bar{t}}),\\
O_3&=&(\hat{q}_+\cdot \hat{p}) (\hat{q}_-\cdot \hat{p}),\\
O_4&=&(\hat{k}_t-\hat{k}_{\bar{t}})\cdot (\hat{q}_+\times \hat{q}_-).
\end{eqnarray*}
where $\hat{q}_+$ ($\hat{q}_-$) is the unit vector of $l^+$ ($l^-$) moving direction in the top quark
 rest frame, $\hat{k}_t$ ($\hat{k}_{\bar t}$) is the unit vector of $t$ ($\bar t$) moving direction 
in the $\tth$ rest frame, and $\hat{p}$ is the unit vector of $\tth$ system moving direction in the 
$pp$ rest frame. Obviously, $O_4$ is a parity violated observable which is sensitive to the parity 
violated interactions. It is proportional to the interference term between scalar and pseudoscalar components.
Therefore it will disappear for the pure scalar or the pure pseudoscalar Higgs.

For the numerical results, we define the expectation value of the operator as
\begin{equation}
<O_i>=\frac{\int O_i  d\sigma}{\int d\sigma}, ~~~~~~~~~(i=1,2,3,4),
\label{eq:average}
\end{equation} 
where $\sigma$ is the cross section of process (\ref{pp2tth}).
In Table~\ref{tab:sm-ob} we display the observables with 
the collision energy of 13 TeV and 33 TeV at the LHC.  
 One can notice that the gluon fusion and quark pair annihilation subprocesses contribute opposite sign for the 
observables in the $\tth$ production, which is the same as in the $t\bar t$ production~\cite{Bernreuther:2004jv}.
\begin{table}[h] 
\begin{center}
\begin{tabular}{cccc|ccc} \hline \hline 
&\multicolumn{3}{c|}{13 TeV}&\multicolumn{3}{c}{33 TeV} \\ \hline
         &  $<0_1>$ &  $<0_2>$ &$<0_3>$  &  $<0_1>$ &  $<0_2>$ &$<0_3>$ \\ \hline
$q\bar q$&-0.0354   & 0.0077   &-0.0362  &-0.0167   & 0.0035   &-0.0169 \\ \hline
$gg$     & 0.0861   &-0.0304   & 0.0172  & 0.0979   &-0.0365   & 0.0177 \\ \hline
total    & 0.0507   &-0.0227   &-0.0190  & 0.0812   &-0.0330   & 0.0008 \\ \hline \hline 
\end{tabular}
\end{center}
\caption{Observables at the LHC 13 TeV and 33 TeV for $m_H=125$ GeV.}\label{tab:sm-ob}
\end{table}
%%%%%%%%%%%%%%%%%%%%%%%%%%%%%%%%%%%%%%%%%%%%%%%%%%%%%%%%%%%%%%%%%%%%%%%%%
\subsection{The Heavy Higgs production associated with $t \bar t$}
\begin{figure}
\centering
\includegraphics[width=0.55\textwidth]{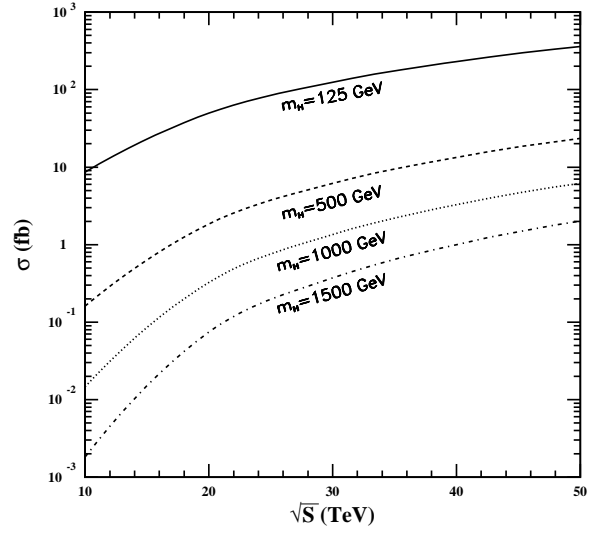} 
\caption{The cross sections for process (\ref{pp2tth}) with respect to the 
center-of-mass energy for different scalar Higgs masses. }
\label{fig:cs-hea}
\end{figure}

%%%%%%%%%%%%%%%%%%%%%%%%%%%%%%%%%%%%%%%%%%%%%%%%%%%%%%%%%%%%%%%%%%%%%%%%%
It is possible for a scalar heavy Higgs production associated with the top quark pair.
From the Lagrangian of (\ref{eq:lag2hdm}), it can be found that the form of 
the heavy Higgs boson coupling to the top quark is the same as the Higgs boson in the SM. 
On the condition of $\epsilon^t_{H_2}=1$, the cross sections of process (\ref{pp2tth}) 
 with respect to different Higgs masses are displayed in Fig.~\ref{fig:cs-hea}. The heavy Higgs properties 
can be investigated at a high collision energy or with a high luminosity.
\begin{figure}
\centering
\includegraphics[width=0.4\textwidth]{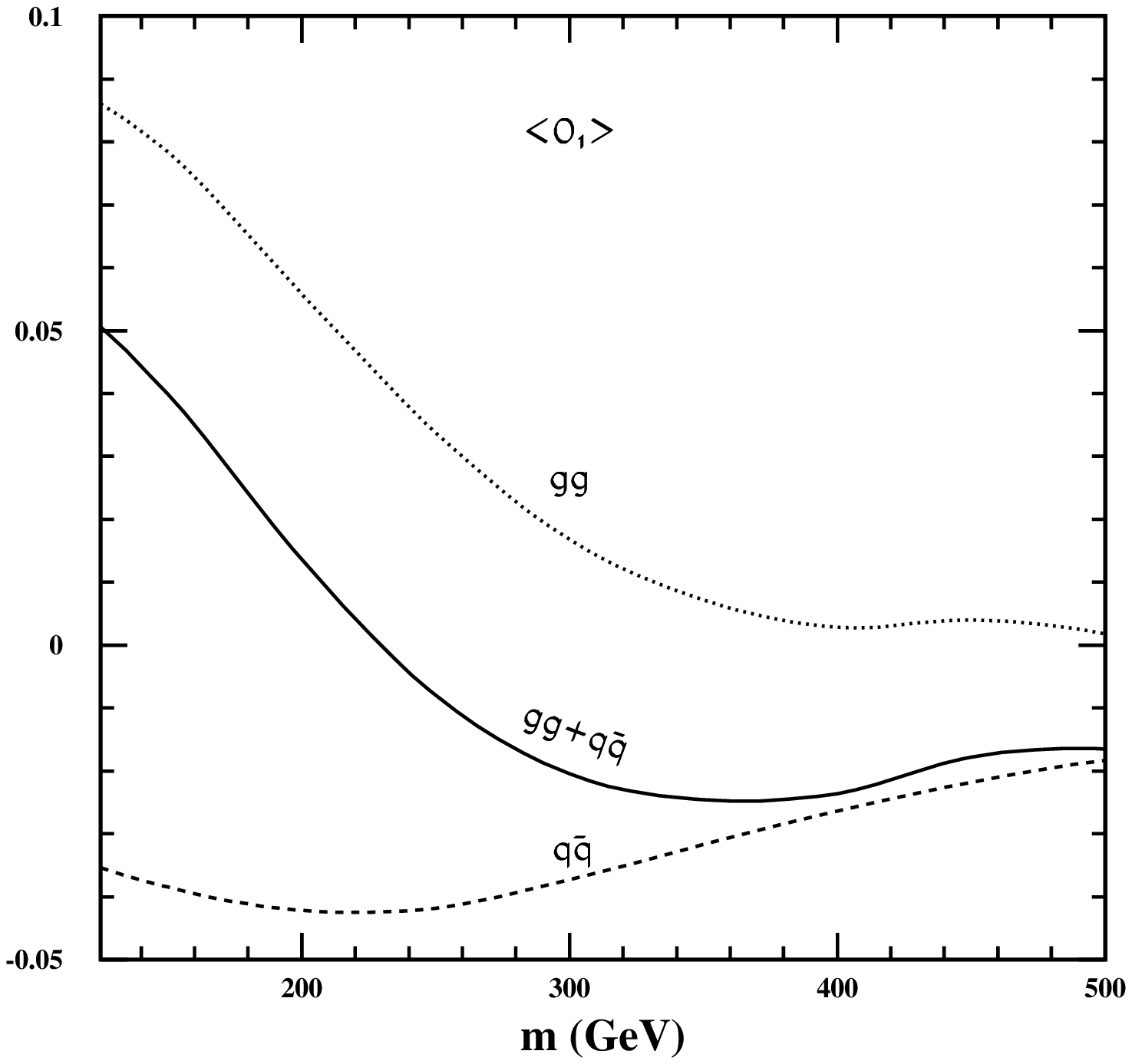} 
\includegraphics[width=0.4\textwidth]{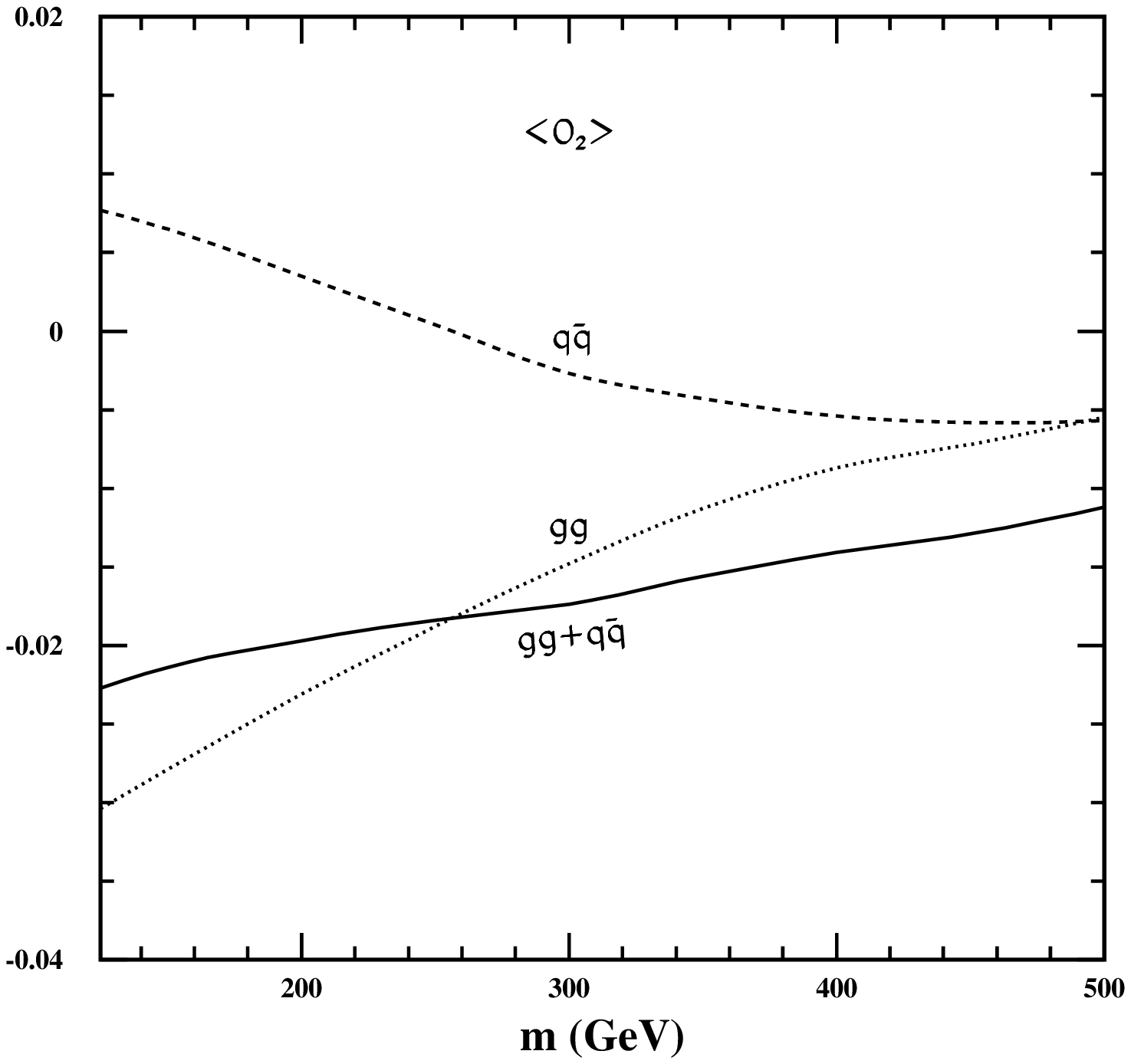}
\includegraphics[width=0.4\textwidth]{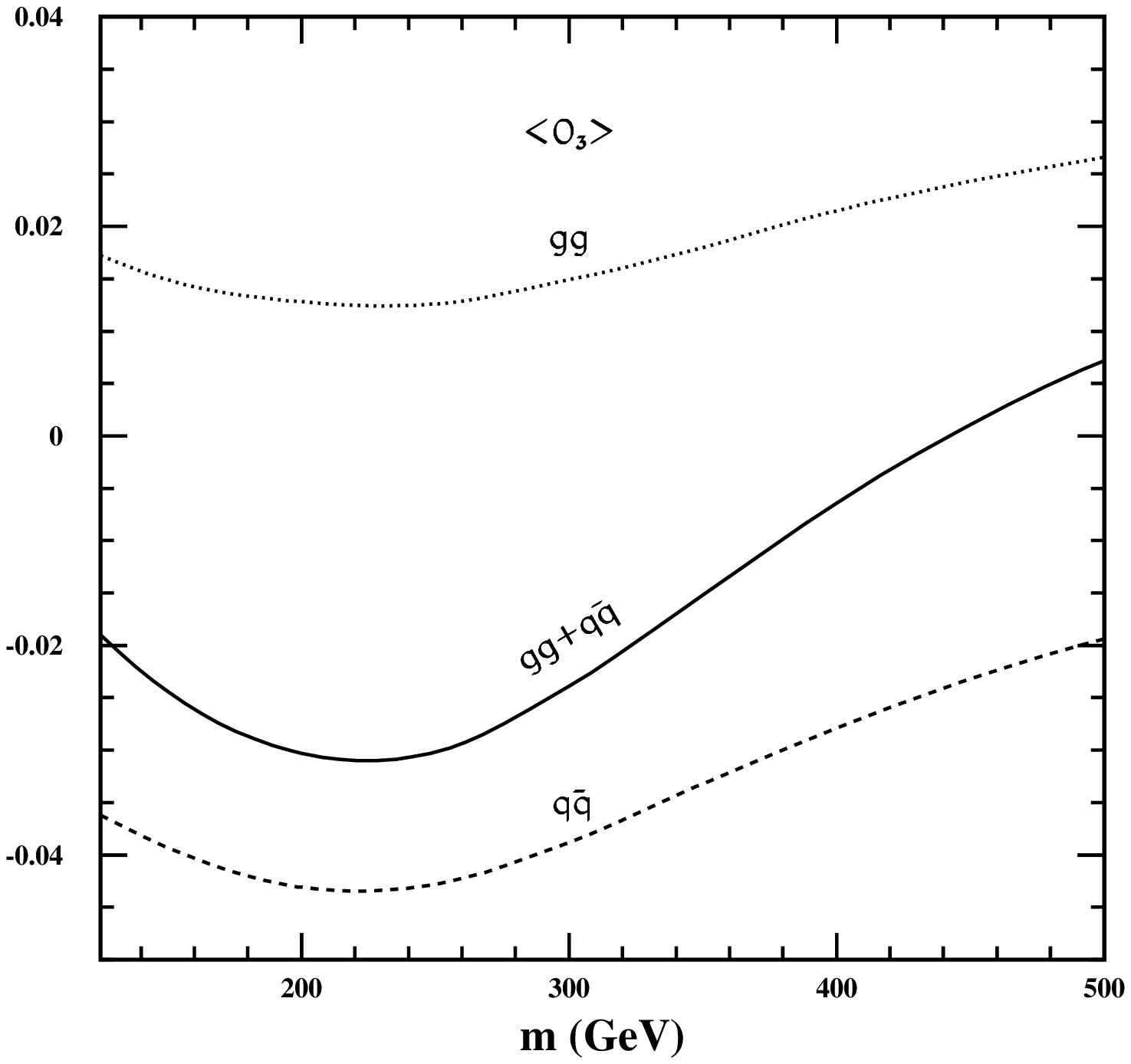}
\caption{The observables of process (\ref{pp2tth}) at 13 TeV LHC with respect to the mass of 
Higgs boson. }
\label{fig:runmass}
\end{figure}

\begin{table}[ht]
\begin{center}
\begin{tabular}{cccc|ccc} \hline \hline
&\multicolumn{3}{c|}{13 TeV}&\multicolumn{3}{c}{33 TeV} \\ \hline
         &  $<0_1>$ &  $<0_2>$ &$<0_3>$ &  $<0_1>$ &  $<0_2>$ &$<0_3>$ \\ \hline
$q\bar q$&  -0.0183 &  -0.0058  &-0.0195&  -0.0059 & -0.0024  &-0.0064 \\ \hline
$gg$     &   0.0036 &  -0.0054  & 0.0267&   0.0090 & -0.0077  & 0.0227 \\ \hline
total    &  -0.0147 &  -0.0112  & 0.0072&   0.0031 & -0.0101  & 0.0163 \\ \hline \hline 
\end{tabular}
\end{center}
\caption{Observables at the LHC 13 TeV and 33 TeV for $m_H=500$ GeV.}\label{tab:hea-13}
\end{table}

We display the contributions from the $gg$ and $q\bar q$ subprocesses for $<O_1>$, $<O_2>$ and $<O_3>$ with 
respect to various Higgs mass at LHC 13 TeV in Fig.~\ref{fig:runmass}. One can notice that for 
$<O_1>$ and $<O_3>$ the contributions of these two subprocesses have different sign. While for $<O_2>$, when the
mass of Higgs is larger than 250 GeV, they have the same sign. So the spin observables are related to
 the Higgs mass. As an example, we list the results of the spin observables in Table~\ref{tab:hea-13} 
with the Higgs mass of 500 GeV at the LHC 13 TeV and 33 TeV.

%%%%%%%%%%%%%%%%%%%%%%%%%%%%%%%%%%%%%%%%%%%%%%%%%%%%%%%%%%%%%%%%%%%%%%%%%
\subsection{The scalar-pseudoscalar mixing Higgs production associated with $t \bar t$}
\begin{figure}
\centering
\includegraphics[width=0.55\textwidth]{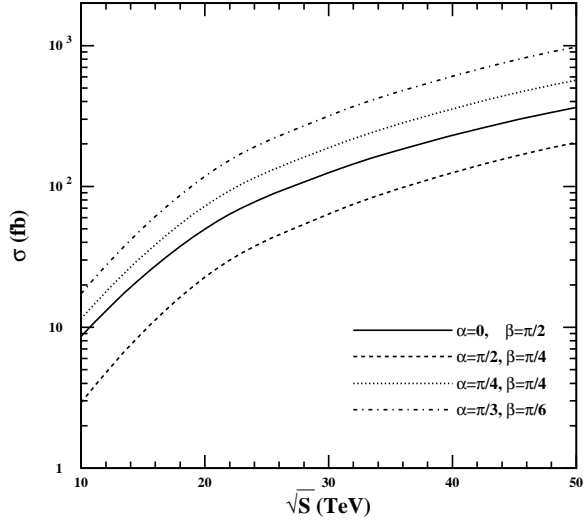}
\caption{The cross sections for process (\ref{pp2tth}) with the center-of-mass 
energy and the Higgs mass 125 GeV.}
\label{fig:cs-cp}
\end{figure}

%%%%%%%%%%%%%%%%%%%%%%%%%%%%%%%%%%%%%%%%%%%%%%%%%%%%%%%%%%%%%%%%%%%%%%%%% 
A toy model can be used to illustrate the CP properties of Higgs boson.
Supposing that the light CP-even and CP-odd Higgs bosons are mass degeneracy in the Two Higgs Doublet Models, 
One can write the interaction between the light Higgs and top quark pair in a general formula as the follows,
\begin{equation}
{\cal L}_{\tth}=-\frac{m_t}{v}(\epsilon^t_{H_1}-i\epsilon^t_{A}\gamma_5)\tth,
\end{equation}
where $\epsilon^t_{H_1}=\cos\alpha/\sin\beta$ and $\epsilon^t_{A}=\cot \beta$ with $\alpha$ the mixing 
angle of the scalar fields. For our numerical results we choose the corresponding values of $\alpha$ and
 $\beta$ as in Table~\ref{tab:para} and $m_H=125$ GeV as an example. $\alpha=0$ and $\beta=\pi /2$ 
corresponds to the SM-Higgs, and $\alpha=\pi/2$ and 
$\beta=\pi /4$ stands for the pseudoscalar Higgs (A). The other two cases stand for the scalar-pseudoscalar
 mixing Higgs. The corresponding cross sections of process (\ref{pp2tth}) at the hadron collider are  
displayed in Fig.~\ref{fig:cs-cp} with different $\tth$ couplings. 
The observables corresponding to process (\ref{pp2tth}) at the LHC are listed in 
Table \ref{tab:cp13tev} and \ref{tab:cp33tev} for the collision energy of 13 and 33 TeV. 
 The values of these observables 
are different in the scalar, pseudoscalar and scalar-pseudoscalar mixing Higgs production. The gluon
 fusion and the quark pair annihilation subprocesses 
contribute the same sign for the observables $<O_2>$ and $<O_3>$ from pseudoscalar Higgs production,
 which differs from the scalar Higgs production. It is found that  $<O_4>$ is a characteristic quantity to distinguish 
the scalar-pseudoscalar mixing Higgs from the pure scalar and the pseudoscalar Higgs. So the precision measurement of 
these observables will be helpful to study the $\tth$ interaction and the properties of Higgs.

\begin{table}[ht]
\begin{center}
\begin{tabular}{ccccc} \hline \hline 
$\alpha$           &   0      &  $\pi$/2   &$\pi$/4   & $\pi$/3  \\ \hline
$\beta$            & $~~\pi$/2 ~~ &  $~~\pi$/4~~   &~~$\pi$/4~~  & ~~$\pi$/6~~ \\ \hline
$\epsilon^t_{H_1}$ & 1        &  0         &   1& 1   \\ \hline
$\epsilon^t_{A}$   & 0        &1           &1         &$\sqrt{3}$ \\ \hline \hline 
\end{tabular}
\end{center}
\caption{Parameters set in the numerical calculations.}\label{tab:para}
\end{table}
\begin{table}[ht]
\begin{center}
\begin{tabular}{ccccc} \hline \hline 
$\alpha$ & $\beta$     & $q\bar q$  & $gg$   & total   \\ \hline 
$\pi$/2  & $\pi$/4     & -0.0037    &0.0092    &0.0055   \\ \hline
$\pi$/4  & $\pi$/4     & -0.0267    &0.0650    &0.0383   \\ \hline
$\pi$/3  & $\pi$/6     & -0.0185    &0.0451    &0.0266 \\ \hline \hline
\end{tabular}~~~~~~~~~~~~~
\begin{tabular}{ccccc} \hline \hline 
$\alpha$ & $\beta$     & $q\bar q$  & $gg$    &total   \\ \hline 
$\pi$/2  & $\pi$/4     & -0.0089   &-0.0436   &-0.0525   \\ \hline
$\pi$/4  & $\pi$/4     &  0.0031   &-0.0340   &-0.0309   \\ \hline
$\pi$/3  & $\pi$/6     & -0.0012   &-0.0374   &-0.0386 \\ \hline \hline
\end{tabular} \\ 
\vspace{2 mm}
(a)~~~~$<O_1>$~~~~~~~~~~~~~~~~~~~~~~~~~~~~~~~~~~~~~~~~~~~~~~~~~~~~ (b)~~~~$<O_2>$\\
\vspace{2 mm}
\begin{tabular}{ccccc} \hline \hline 
$\alpha$ & $\beta$     & $q\bar q$  & $gg$   & total   \\ \hline 
$\pi$/2  & $\pi$/4     &  0.0025    &0.0138   & 0.0163   \\ \hline
$\pi$/4  & $\pi$/4     & -0.0253    &0.0160   &-0.0093   \\ \hline
$\pi$/3  & $\pi$/6     & -0.0153    &0.0153   &0         \\ \hline \hline
\end{tabular}~~~~~~~~~~~~~
\begin{tabular}{ccccc} \hline \hline 
$\alpha$ & $\beta$     & $q\bar q$  & $gg$    & total   \\ \hline 
$\pi$/2  & $\pi$/4     & 0          &0        &0   \\ \hline
$\pi$/4  & $\pi$/4     & -0.0104    &0.0437   &0.0333   \\ \hline
$\pi$/3  & $\pi$/6     & -0.0116    &0.0484    &0.0368 \\ \hline \hline
\end{tabular} \\
\vspace{2 mm}
(c)~~~~$<O_3>$~~~~~~~~~~~~~~~~~~~~~~~~~~~~~~~~~~~~~~~~~~~~~~~~~~~~ (d)~~~~$<O_4>$\\
\end{center}
\caption{Observables at the LHC 13 TeV for the mass of Higgs 125 GeV.}
\label{tab:cp13tev}
\end{table}
\begin{table}[ht]
\begin{center}
\begin{tabular}{ccccc} \hline \hline 
$\alpha$ & $\beta$     & $q\bar q$  & $gg$   & total   \\ \hline 
$\pi$/2  & $\pi$/4     & -0.0013    &0.0097    &0.0084   \\ \hline
$\pi$/4  & $\pi$/4     & -0.0114    &0.0674    &0.0560   \\ \hline
$\pi$/3  & $\pi$/6     & -0.0073    &0.0439    &0.0366 \\ \hline \hline
\end{tabular}~~~~~~~~~~~~~
\begin{tabular}{ccccc} \hline \hline 
$\alpha$ & $\beta$     & $q\bar q$  & $gg$    &total   \\ \hline 
$\pi$/2  & $\pi$/4     & -0.0035   &-0.0445   &-0.0480   \\ \hline
$\pi$/4  & $\pi$/4     &  0.0011   &-0.0392   &-0.0381   \\ \hline
$\pi$/3  & $\pi$/6     & -0.0008   &-0.0414   &-0.0422 \\ \hline \hline
\end{tabular} \\ 
\vspace{2 mm}
(a)~~~~$<O_1>$~~~~~~~~~~~~~~~~~~~~~~~~~~~~~~~~~~~~~~~~~~~~~~~~~~~~ (b)~~~~$<O_2>$\\
\vspace{2 mm}
\begin{tabular}{ccccc} \hline \hline 
$\alpha$ & $\beta$     & $q\bar q$  & $gg$   & total   \\ \hline 
$\pi$/2  & $\pi$/4     &  0.0010    &0.0127    &0.0137  \\ \hline
$\pi$/4  & $\pi$/4     & -0.0107    &0.0159    &0.0052   \\ \hline
$\pi$/3  & $\pi$/6     & -0.0060    &0.0146    &0.0086 \\ \hline \hline
\end{tabular}~~~~~~~~~~~~~
\begin{tabular}{ccccc} \hline \hline 
$\alpha$ & $\beta$     & $q\bar q$  & $gg$    & total   \\ \hline 
$\pi$/2  & $\pi$/4     & 0          &0        &0   \\ \hline
$\pi$/4  & $\pi$/4     & -0.0045    &0.0498   &0.0453   \\ \hline
$\pi$/3  & $\pi$/6     & -0.0046    &0.0510   &0.0464 \\ \hline \hline
\end{tabular} \\
\vspace{2 mm}
(c)~~~~$<O_3>$~~~~~~~~~~~~~~~~~~~~~~~~~~~~~~~~~~~~~~~~~~~~~~~~~~~~ (d)~~~~$<O_4>$\\
\end{center}
\caption{Observables at the LHC 33 TeV for the mass of Higgs 125 GeV.}
\label{tab:cp33tev}
\end{table}

%%%%%%%%%%%%%%%%%%%%%%%%%%%%%%%%%%%%%%%%%%%%%%%%%%%%%%%%%%%%%%%%%%%%%%%%%
\subsection{The signal and backgrounds at the LHC}
%%%%%%%%%%%%%%%%%%%%%%%%%%%%%%%%%%%%%%%%%%%%%%%%%%%%%%%%%%%%%%%%%%%%%%%%% 
To get an idea of the sensitivity one absolutely needs to include backgrounds. For the signal of
$p p \to t \bar t H \to b \bar b l^+ l^- \nu_l \bar {\nu_l} + b \bar b$, the detector signal 
for our investigated process will be two leptons, four b-jets and missing energy. 
The main backgrounds with the same collider signal are 
\begin{eqnarray}
pp &\to& t \bar t j j \nonumber \\
pp &\to& t \bar t b \bar b \\
pp &\to& t \bar t Z \to t \bar t b \bar b, \nonumber
\end{eqnarray}
where $j$ stands for the light jet. According to the analysis
 in reference~\cite{Aad:2015jfa}, the top quark can be reconstructed by solving the kinematic 
equations obtained when imposing energy-momentum conservation at each of the decay vertices of 
the process. From the leptonic decay channel, the top quark reconstruction efficiency can be up 
to 80\%. The reconstruction details can be found in~\cite{Santos:2015dja}. 

For the aim of highlighting the signal process from the backgrounds, we set the kinematics cuts as
\begin{equation}
\left\{ 
\begin{array} {c}
P_T \ge 20~ GeV\\
~|y|\le 3.0~~~~~~~~~,
\end{array}
 \right.
\end{equation} 
where the $P_T$ is the transverse momentum of the charge leptons and the b-jets, and $y$ is the corresponding rapidity.  
The differences of the signal and the background processes are mostly 
from the two jets which do not derive from the top quarks, thus we adopt the invariant mass of these two 
jets which is close to  the Higgs mass, i.e.,
\begin{equation}
|M_{jj}-M_H|\le0.1M_H.
\end{equation} 
We simulated the backgrounds processes by the MADGRAPH programs~\cite{Alwall:2014hca}, where a sets of acceptant cuts are adopted. 
The results are summarized in Table~\ref{tab:bg}.
One can notice that the backgrounds are more tremendous than the signal process before cuts, while the three orders of magnitude
gaps are disappear after the cuts. For the SM-like Higgs boson production at LHC 13 TeV, the significance can be up to
 $S/B=0.3$ and $S/\sqrt{B}=10.1$ for the integrated luminosity of $300 fb^{-1}$. When the collision energy is up to 33 TeV, 
the heavy Higgs with a mass of 500 GeV can be detected with a significance of $S/B=0.07$ and $S/\sqrt{B}=5.79$ for the 
integrated luminosity of $1000 fb^{-1}$.

\begin{table}[h] 
\begin{center}
\begin{tabular}{|c|c|c|c|c|c|c|c|}  \hline 
\multirow{2}{*}{$\sqrt{S}$}& \multirow{2}{*}{$m_H$ (GeV)}&\multicolumn{2}{c|}{before cut}&
\multicolumn{4}{c|}{after cut}  \\ 
\cline{3-8}&&~~~~~signal~~~~~& backgrounds &~~~~~signal~~~~~~
& backgrounds  &~~~$S/B$~~~&~$S/\sqrt{B}$~ \\ \hline
\multirow{2}{*}{13 TeV}&125&2.05& \multirow{2}{*}{8039.7}&1.14& 3.81& 0.30& 10.1\\  
\cline{2-3}\cline{5-8}&500&0.051&&0.031&0.43&0.07& 0.82\\ \hline  
\multirow{2}{*}{33 TeV}&125&18.02&\multirow{2}{*}{$1.24 \times 10^5$}&9.86&44.2&0.22& 46.9\\ 
\cline{2-3}\cline{5-8}&500&0.65&&0.49&7.16&0.07&5.79 \\ \hline 

\end{tabular}
\end{center}
\caption{Summary of the cross sections for the signal and background processes at the LHC 13 TeV and 33 TeV 
before and after cuts. The significances are listed in the last two rows.}\label{tab:bg}
\end{table}

%%%%%%%%%%%%%%%%%%%%%%%%%%%%%%%%%%%%%%%%%%%%%%%%%%%%%%%%%%%%%%%%%%%%%%%%%
\section{Summary}\label{summary}
%%%%%%%%%%%%%%%%%%%%%%%%%%%%%%%%%%%%%%%%%%%%%%%%%%%%%%%%%%%%%%%%%%%%%%%%%
A large number of Higgs events will be accumulated at the LHC, and the properties 
of Higgs boson should be addressed. In the SM, one CP-even Higgs boson is assumed 
via the simplest scalar doublet, which has been naturally extended in the new 
physics models, such as the Two Higgs Doublet Models. The phenomenology of Higgs sector 
is extremely rich, since it contains more than one Higgs. The interactions of Higgs boson
coupling to other particles are more complex than those in the SM. Therefore, to discriminate
the new physics models, it is important to study the properties of Higgs. For this aim,
in this paper we study the $\tth$ production and its related spin effects at the LHC.
The top quark spin correlation, reflected by the motion of the 
particles decaying from the top quark pair, is related to the dynamics of $\tth$ production.
 These spin correlations are related to the couplings of the $\tth$ interaction and the Higgs mass. 
To study these spin effects, we adopt the 
observables $<O_i>$ ($i=1,2,3,4$) to investigate the properties of the scalar, pseudoscalar and scalar-pseudoscalar 
mixing Higgs in the $\tth$ production. 
 With the large statistic of $\tth$ events at the LHC,
 the Higgs properties can be clarified so that the new physics models can be discriminated.   
%%%%%%%%%%%%%%%%%%%%%%%%%%%%%%%%%%%%%%%%%%%%%%%%%%%%%%%% %%%%%%%%%%%%%%%%%
\begin{acknowledgments}
%%%%%%%%%%%%%%%%%%%%%%%%%%%%%%%%%%%%%%%%%%%%%%%%%%%%%%%%%%%%%%%%%%%%%%%%%
This work is supported in part by the NSFC with Grant Nos. 11275114, 11325525 and 
11305075 and NSF of Shandong Province with Grant No. ZR2013AQ006.
\end{acknowledgments}
%%%%%%%%%%%%%%%%%%%%%%%%%%%%%%%%%%%%%%%%%%%%%%%%%%%%%%%%%%%%%%%%%%%%%%%%%

\end{document}